\documentclass[twocolumn,twoside,slac_two]{revtex4}
\usepackage{graphicx}
\usepackage{fancyhdr}
\usepackage{graphics}
\usepackage{epstopdf}
\usepackage{textpos}
\usepackage{hyperref}
\newcommand{\ICRC}{32nd ICRC, Beijing, China, 2011}
\begin{document}

\title{\centering Ultra High Energy Cosmic Rays}
\author{
\centering
\begin{center}
Miguel Mostafa
\end{center}}
\affiliation{\centering Colorado State University, Fort Collins; CO 80525; USA}
\begin{abstract}
Ultra high energy particles arrive at Earth constantly.
They provide a beam at energies higher than any man-made accelerator, but at a very low rate.
Two large experiments, the Pierre Auger Observatory and the Telescope Array experiment, have been taking data for several years now covering together the whole sky.
I summarize the most recent measurements from both experiments,
I compare their results and, for a change, I highlight their agreements.
\end{abstract}
\maketitle
\thispagestyle{fancy}

\section{Introduction}
Ultra high energy cosmic rays (UHECRs) are subatomic particles that reach the Earth with macroscopic energies (UHE~$>10^{18}$~eV).
Obviously, accelerating particles to energies of the order of $10^6$~TeV is not easy, and the flux of UHECRs is extremely low:
only a few particles per year per km$^2$ reach us.
Thus, experiments above the Earth's atmosphere are not feasible.
Instead, UHECRs are observed via their interactions with the Earth's atmosphere.
When the primary particle enters the top of the atmosphere, it produces a cascade of secondary particles, a.k.a.\ an extensive air shower.
In the process, an enormous amount of energy is deposited in the atmosphere, and billions of secondary particles reach the ground. 
One detection technique consist of particle counters on the ground to measure the secondary particles, 
and to infer the properties of the primary particle.
In collider terms, UHECRs provide beams of particles with energies of millions of TeV, our experiments are fixed target experiments, and the atmosphere is our calorimeter.
The center-of-mass of the first interactions is of the order of hundreds of TeV.

Another detection technique uses the fact that part of the energy deposited in the atmosphere by UHECRs is re-emitted isotropically as ultra-violet light.
The process is very inefficient, and a primary particle of $10^{20}$~eV is seen from the ground as a 50~W light bulb crossing the atmosphere at the speed of light.

There are currently two large experiments taking statistically significant data at ultra-high energies.
Namely,  the Pierre Auger Observatory (Auger) in Argentina, and the Telescope Array experiment (TA) in Utah, USA.
In this paper I present the most recent measurements produced by these two large International collaborations.
Most of these results have been presented recently at the International Cosmic Ray Conference in Beijing, China. 

\section{Detectors of UHECRs}

A powerful feature of the  design of both Auger and TA is the capability of observing
air showers simultaneously by the two different but complementary
techniques mentioned in the Introduction.

Both experiments consist of a large array of particle detectors on the ground.
This surface array measures the particle
densities as the shower strikes the Earth. 
Due to its large area and almost 100\% duty cycle, 
it provides the statistics needed to study these rare particles.

On dark moonless nights, air fluorescence\footnote{As in most of particle physics, and physics in general, the word fluorescence is a misnomer, but it has been used even before I was born, 
so I go here with the commonly used naming convention.} telescopes
record the development of  
the 
shower that results from the interaction of the primary particle with
the upper atmosphere. 
By recording the light produced by the developing air
shower, fluorescence telescopes can make a nearly calorimetric
measurement of the energy.  
This energy calibration can then be
transferred to the surface array with its 100\% duty factor and large
event gathering power.  
The energy 
determination 
is therefore
done with minimal reliance on
either numerical simulations, 
or assumptions about the composition, 
or interaction models.
(But as anyone that has ever participated in the energy calibration of a calorimeter can understand, the calibration of a time dependent calorimeter of thousands of cubic kilometers is not a trivial task.)

\subsection{The Pierre Auger Observatory}

The Pierre Auger Observatory features an array of over 1600 water Cherenkov
 detectors spread over 3000~km$^2$, and arranged on a triangular
grid, with the sides of the triangles being 1.5~km~\cite{Auger}.  
Four fluorescence detector (FD) stations, each
containing six fixed telescopes designed to detect air-fluorescence
light, overlook the ground array.  
The surface detector (SD) stations
measure the density distribution of the air shower cascade as it
strikes the ground while the FD telescopes measure the
light produced by atmospheric nitrogen excited by the cascading
shower. 
This combined approach is called the \textit{hybrid} detection technique.

The primary purpose of the FD is to measure the
longitudinal profile of showers recorded by the SD whenever it is dark
and clear enough to make reliable measurements of atmospheric
fluorescence from air showers.  
The integral of the longitudinal profile is used to determine the shower energy, 
and the speed of shower development is indicative of the primary particle's mass.  
The hybrid detector has better angular resolution than the surface array alone.

\begin{figure}[!ht]
\includegraphics[width=0.45\textwidth]{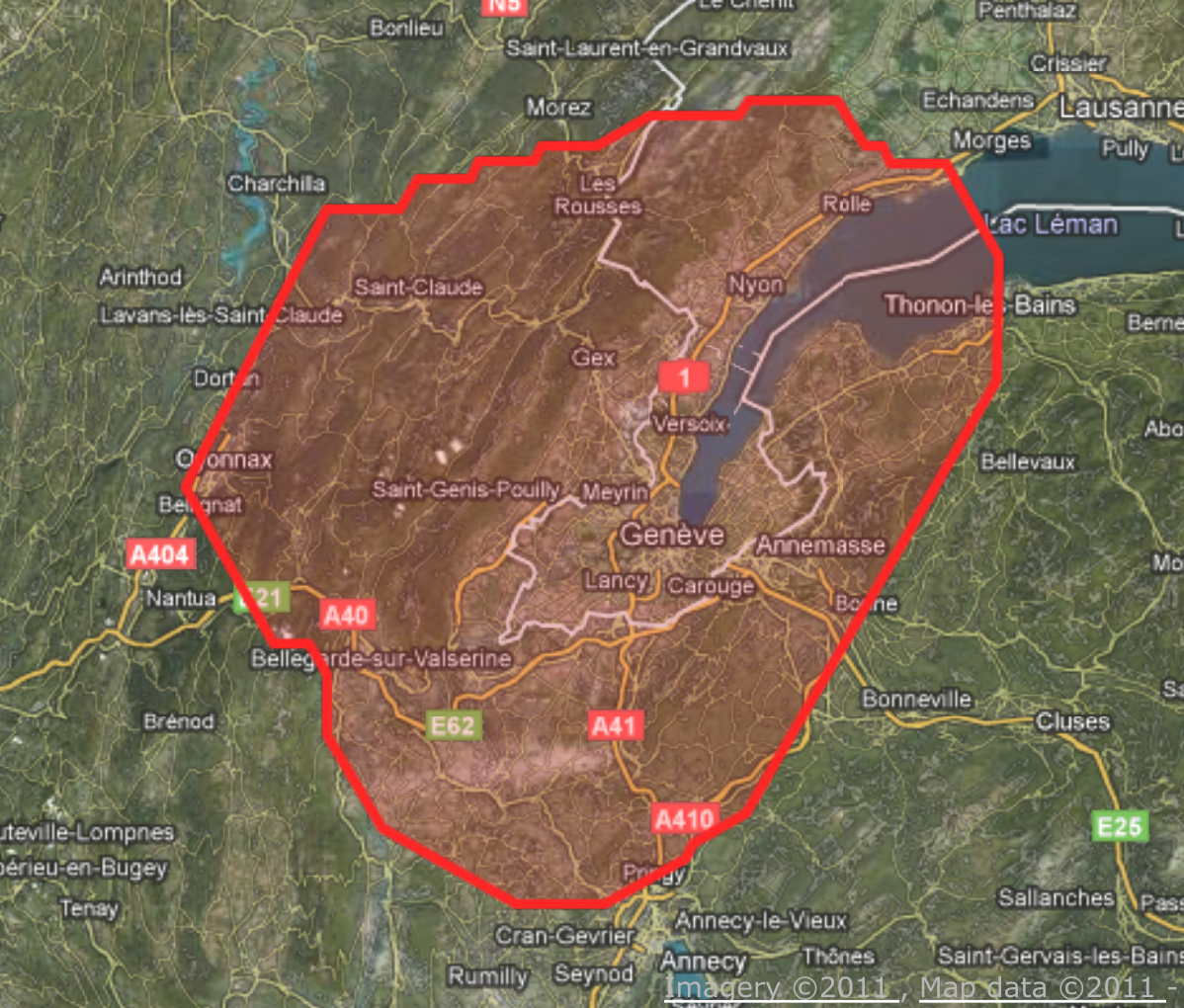}%
\caption{Comparison between the size of the Auger array (red contour) and the area around Geneva, Switzerland. 
[Image produced using the Pierre Auger Public Event Display, available at
\url{http://auger.colostate.edu/ED/}.}
\label{fig:augerarray}
\end{figure}

The Auger site in Argentina is in the Province of Mendoza near the city of
Malarg\"{u}e. 
The site is located at a latitude of $35^\circ$ south with a mean altitude of 1400~m a.s.l.  
The site is a relatively flat alluvial plain, sufficiently large to easily encompass the required 3000~km$^2$ footprint of the array.  
In Fig.~\ref{fig:augerarray} I compare the size of the Auger array with the area around the LHC.
There are convenient elevated positions on the edge of the array that allow placement of
the four FD stations slightly above ground level.  


%


Each SD station is a cylindrical water tank of $3.7$~m in diameter and $1.2$~m deep, holding 12~metric tons of filtered highly deionized water. 
Because of the height and volume of the detectors, they are nearly uniformly sensitive to particle track length out to very large zenith angles. 
The water is enclosed within a bag with an inner layer made of white Tyvek\textsuperscript{\textregistered}, which reflects light diffusely. 
Three  (downward-looking) $9"$ photomultiplier tubes detect the Cherenkov light.

Each FD site contains six identical telescopes. 
Ultra-violet light enters the telescope window through a $1.1$~m radius diaphragm, 
and is collected by a $3.5\times3.5$~m spherical mirror, and focused onto a photo-multiplier (PMT) camera. 
Each camera contains 440 hexagonal 45~mm PMTs. 
Each PMT covers a $1.5^\circ$ diameter region of the sky. 
The field of view of a single telescope covers $30^\circ$ in azimuth and $28.6^\circ$ in elevation. 
An optical filter matched to the fluorescence spectrum (approximately $300$~nm to $400$~nm) is placed over the telescope diaphragm to reduce night-sky noise. 
In addition, the diaphragm contains an annular corrector lens as part of the Schmidt telescope design. 
The effect of the lens is to allow an increase in the radius of the telescope diaphragm from $0.85$~m to $1.1$~m (increasing the effective light collecting area by a factor of two) while maintaining an optical spot size of $0.5^\circ$.

The Auger
Observatory started collecting data (with over 200 operational SD stations) in January 2004. 
The construction of the Observatory was completed (over 1600 operational SD stations, and 24 operational FD telescopes) in June, 2008.
Data from the SD have been taken since then with a duty cycle of almost 100\%.
Both surface and fluorescence detectors have been running simultaneously 14\% of the time. 
The resolution for the arrival direction of cosmic rays is between $1^\circ$ and $2^\circ$ (depending on the energy) for the SD only, and it is less than $1^\circ$ in hybrid mode.
The energy resolution of Auger is $\sim15\%$ (for cosmic rays with energy above 3~EeV), and there is a systematic uncertainty of $\sim22\%$ on the overall energy scale.

\subsection{The Telescope Array Experiment}
The Telescope Array experiment~\cite{dpf2011_cchjui:TA} 
combines the large area scintillation ground array technique developed by 
the Akeno Giant Air Shower Array (AGASA)~\cite{AGASA} in Japan
with that of the air fluorescence method pioneered by the Fly's Eye~\cite{FlysEye} at the University of Utah, 
and later improved by
the High Resolution Fly's Eye (HiRes) experiment~\cite{HiRes}.

The TA experiment is located in the central western desert of Utah, near the city of Delta, and 
it has three FD stations located at the periphery of a ground array of 507 SD units.
Each SD unit consists of a scintillation counter mounted on a raised steel frame.  
They are deployed in a square grid of 1.2~km nearest-neighbor spacing, and the full array covers a total of about 730~km$^{2}$.

Each of the three FD stations has a field of view of about $30^{\circ}$ in elevation and about 110$^{\circ}$ in azimuth.  
A total of 38 fluorescence telescopes are divided into three stations.  
One station is a recycled HiRes detector.
The other two FD stations were built in Japan 
based on essentially the same specifications as the telescopes used by the HiRes experiment, 
but with larger mirrors of 6.8~m$^{2}$ area (compared to 5.2~m$^{2}$ for HiRes).  
Each site consists of 12 telescopes with 256-pixel ($16\times16$ in a triangular lattice) cameras.  
Each pixel, instrumented by a hexagonal PMT, covers a cone of 1.1$^{\circ}$ in the sky.

Each TA scintillation counter contains two slabs of double-layered plastic
scintillators with an overall collection area of 3~m$^{2}$.  
The scintillation photons are collected by wavelength-shifting optical fibers laid in extruded grooves on the surface of the scintillators.  
The light collected from the top and bottom layers are each separately collected into a single PMT.  
%
TA started operation in May, 2008. 
The angular resolution of TA is $1.5^\circ$ (above $10^{19}$~eV), the energy resolution is $\sim20\%$,
and there is a systematic uncertainty of $\sim21\%$ on the overall energy scale.

\section{Latest results}

The three main questions about UHECRs relate to
the acceleration mechanism(s) capable of reaching ultra-high energies,
the location of these amazing accelerators,
and the type of particles being accelerated.
To provide answers to these questions we  measure  the energy distribution of UHECRs, their arrival directions, and the air shower characteristics that allow us to infer their identity.

In the following sections I summarize the most recent results from the Auger and TA experiments in these three areas of interest.
\subsection{Energy}

To understand the acceleration mechanism(s), the energy distribution of UHECRs is measured, and then compared to the predictions from different acceleration models. 

The energy range accessible to both Auger and TA covers two interesting spectral features.
The so-called ``ankle" at $\sim4\times10^{18}$~eV where the spectrum hardens.
If the maximum energy achievable by acceleration in relativistic shock waves were high enough to produce galactic cosmic rays up to the ankle,
this feature could be the transition from a steep galactic spectrum to a flat spectrum of extragalactic cosmic rays.
But if the galactic component of cosmic rays ended at lower energy ($\sim2\times10^{17}$~eV),
the ankle could be due to the combination of adiabatic losses 
and $e^\pm$ pair production.

\begin{figure*}[!ht]
\centering
\includegraphics[width=\textwidth]{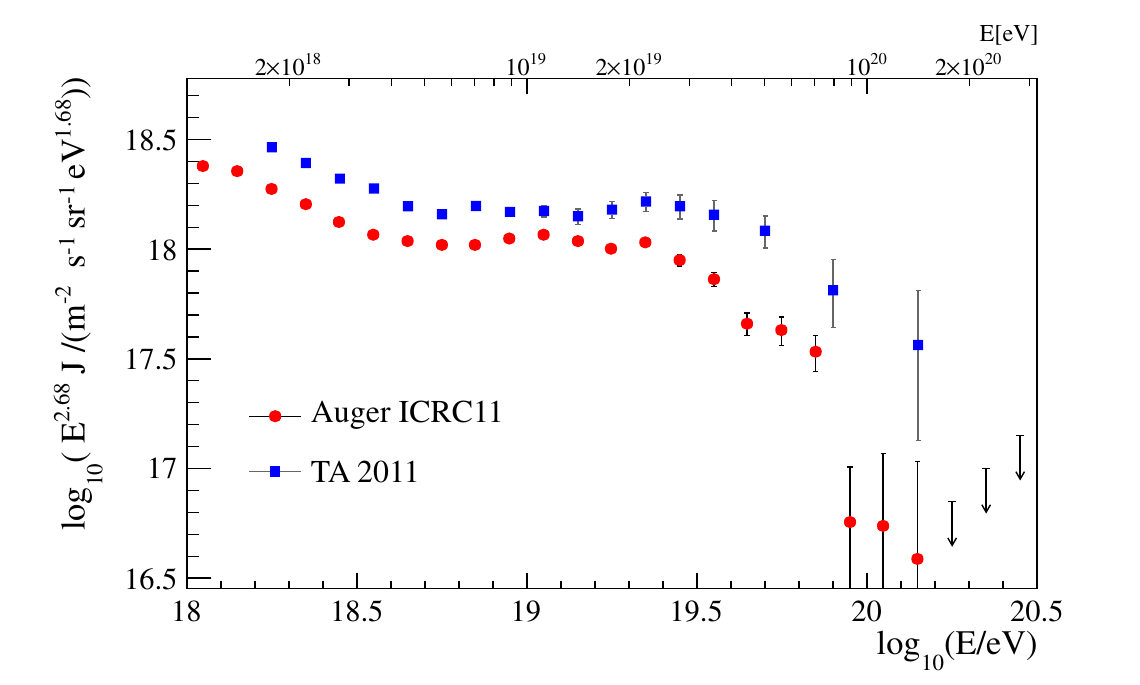}%
\caption{Comparison of the differential UHECR fluxes (multiplied by $E^{2.68}$) measured with the Pierre Auger and Telescope Array surface detectors~\cite{Unger}.
Despite the different energy scale, the spectral features are in good agreement. (See text and table.)} 
\label{fig:spectrum}
\end{figure*}

Another crucial issue is the measurement of the spectrum at energies above  $\sim4\times10^{19}$~eV where the 
spectrum of UHECRs is characterized by a strong flux suppression (the GZK feature) due to 
the interactions between extragalactic cosmic rays and the cosmic background radiation.

The energy spectra measured by Auger and TA are shown in Fig.~\ref{fig:spectrum}.
There is a clear difference in the energy scale.
(This could be compared to a different luminosity normalization for collider experiments.)
The TA spectrum agrees very well with the spectrum previously measured by HiRes by construction.
The TA collaboration finds a difference of 27\% between their SD and the FD that is used to determine the absolute energy scale~\cite{Stokes}.
Despite the different energy scale, 
the key spectral features are in good agreement. 
The summary of these main features is shown in Table~\ref{tab:spectrum}.

\begin{table}[!ht]
\begin{center}
\caption{Spectral features in the UHECR fluxes measured with the Telescope Array experiment~\cite{Stokes} and the Pierre Auger Observatory~\cite{Salamida},
where $\gamma_1$, $\gamma_3$, and $\gamma_2$ are the spectral indices below the ankle, above the GZK suppression, and in between, respectively;
and $E_1$ and $E_2$ are the energies at which they find the ankle and the flux suppression, respectively.%
\label{tab:spectrum}}
   \begin{tabular}{lcc}
                &    \textbf{TA}          & \textbf{Auger} \\\hline
     $\gamma_1$ &  $3.33\pm0.04$ & $3.27\pm 0.02$\\
     $\gamma_2$ &  $2.68\pm0.04$ & $2.68\pm 0.01$\\
     $\gamma_3$ &  $4.2\pm 0.7$  & $4.2\pm 0.1$ \\
     $\log(E_1/\mbox{eV})$ &  $18.69\pm 0.03$  & $18.61\pm 0.01$ \\
     $\log(E_2/\mbox{eV})$ &  $19.68\pm 0.09$  & $19.41\pm 0.02$\\\hline
   \end{tabular}
\end{center}
\end{table}

\subsection{Composition}

Inferring the identity of these particles is the most difficult measurement because UHECRs are not directly detected.
We measure the secondary particles that reach the ground, and the light produced by these secondaries while transversing the atmosphere.
Consequently, the composition is inferred from observables of the extensive air showers.
For experiments with fluorescence telescopes, 
the observable most sensitive to the mass of the primary cosmic ray is the atmospheric depth at which the maximum number of secondary particles is reached.
(The number of secondary particles is proportional to the amount of energy deposited in the atmosphere, which in turn is proportional to the amount of UV light that reaches the telescope.)
The average depth of shower maximum, $X_{\normalsize max}$, depends on both the energy and the mass of the primary particle:
$$\left< X_{max}\right> \propto D_{10} \log(E/A),$$
where $E$ and $A$ are the energy and mass of the primary particle, and $D_{10}$ is known as the elongation rate.
Basically, the more energetic and lighter the cosmic ray, the more penetrating the produced air shower.

\begin{figure*}[!ht]
\centering
\includegraphics[width=0.5\linewidth]{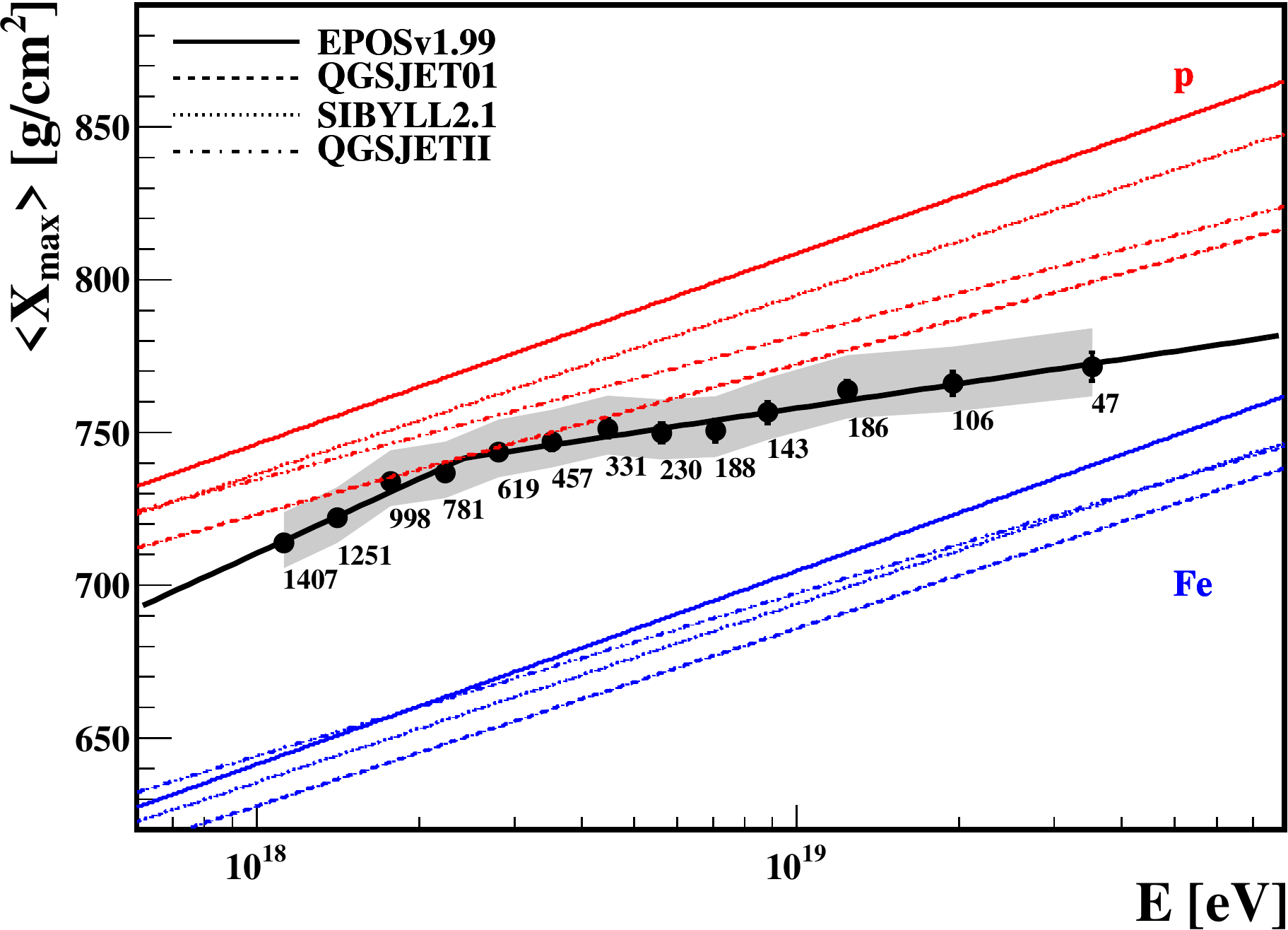}%
\includegraphics[width=0.5\linewidth]{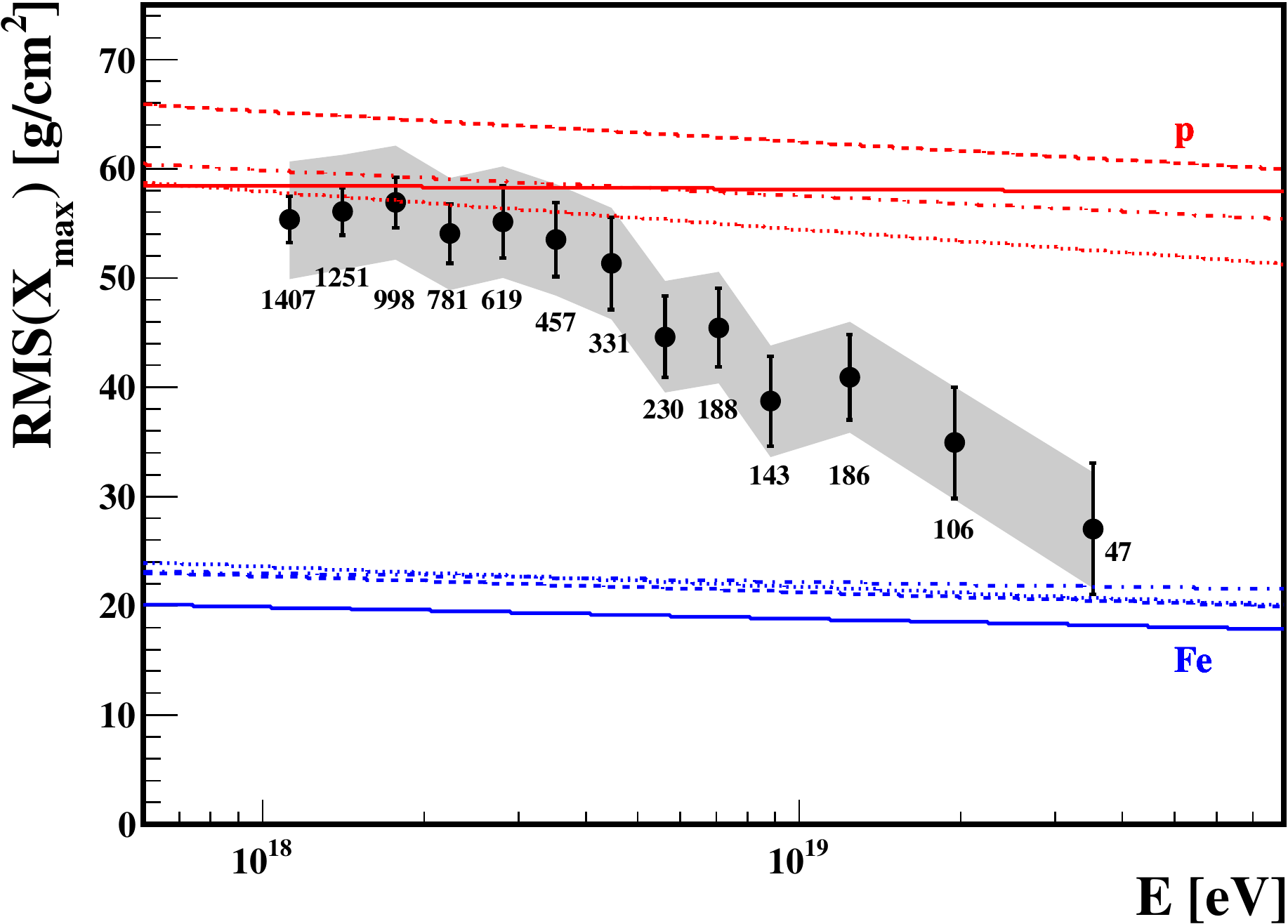}
\caption{Average value of the atmospheric depth at which the longitudinal development of air showers reaches its maximum 
(left) and 
the magnitude of the shower-to-shower fluctuations of the depth of maximum
(right)
as a function of energy measured with the Pierre Auger Observatory~\cite{Facal}.
Data (points) are unfolded from detector effects, 
and directly compared with air shower simulations~\cite{simulations} for proton and iron primaries using different hadronic interaction models~\cite{models}. 
The number of events in each bin is indicated. Systematic uncertainties are indicated as a band.
} 
\label{fig:AugerER}
\end{figure*}

\begin{figure*}[!ht]
\centering
\includegraphics[width=0.5\linewidth]{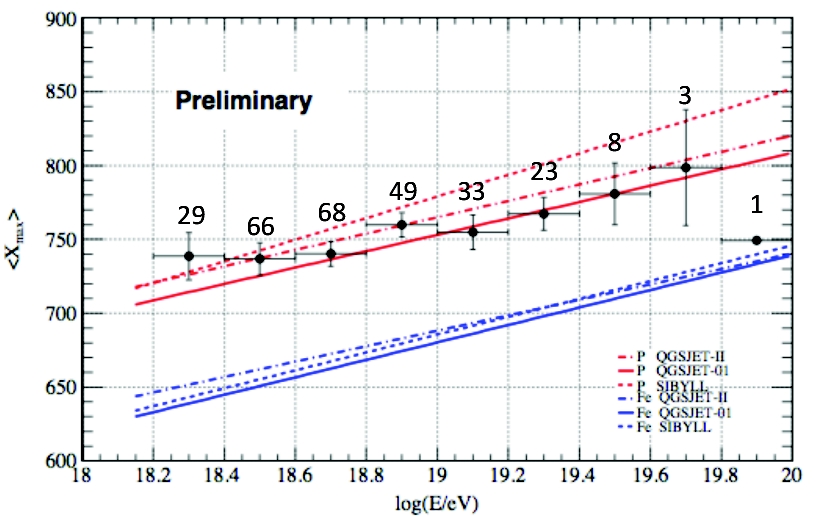}%
\includegraphics[width=0.52\linewidth]{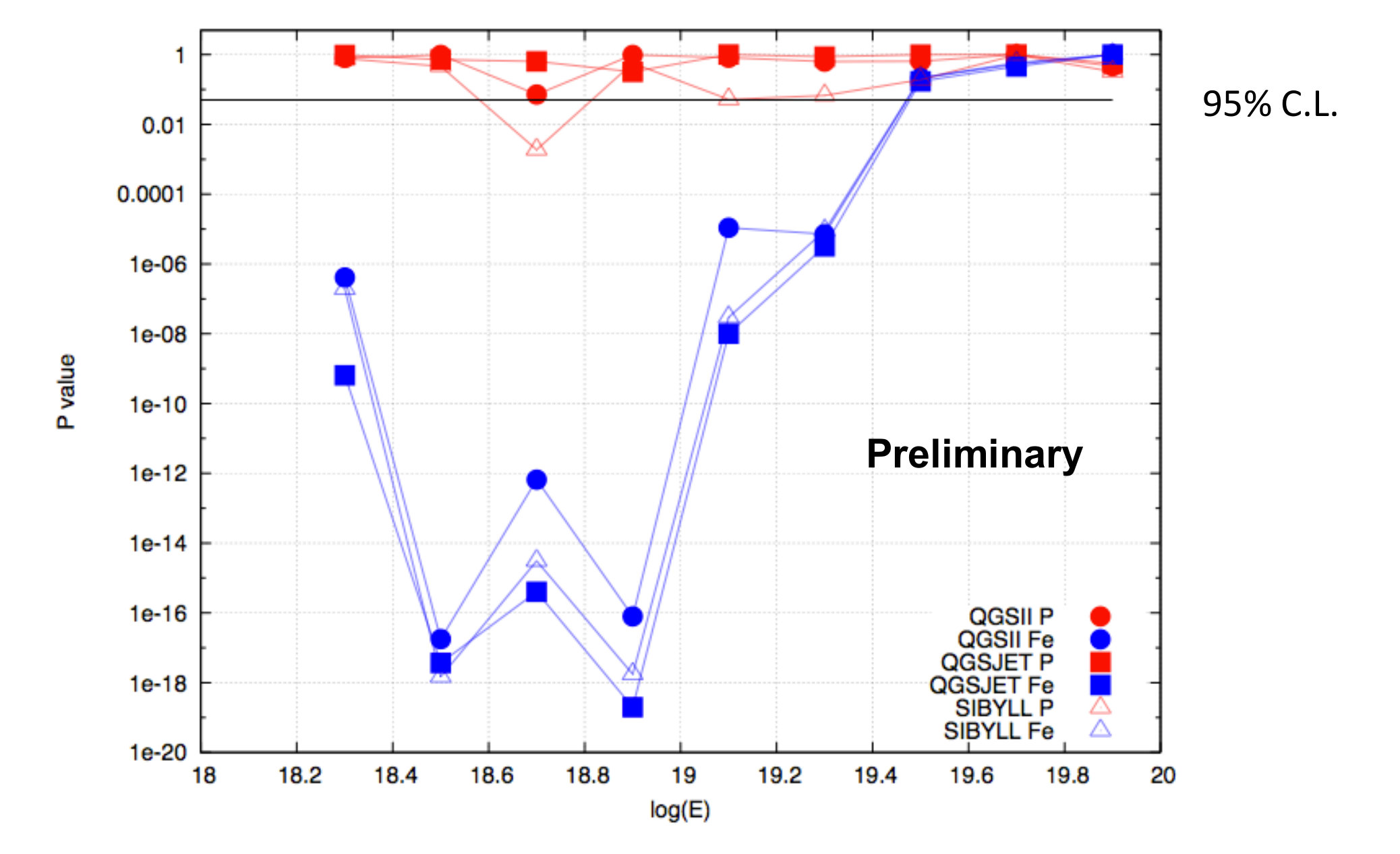}
\caption{{\it Left:} Average value of the atmospheric depth at which the longitudinal development of air showers reaches its maximum 
as a function of energy measured with the Telescope Array experiment~\cite{Tameda}.
Data (points) are compared with  simulations smeared for detector resolution effects.
The number of events in each bin is indicated. 
{\it Right:} Kolmogorov-Smirnov test between the data distributions and the smeared simulations for both proton (red) and iron (blue) primaries.
} 
\label{fig:TAER}
\end{figure*}

The most recent measurement from Auger~\cite{Facal} (TA~\cite{Tameda}) of $\left< X_{max}\right>$ as a function of energy is shown on the left panel of Fig.~\ref{fig:AugerER} (Fig.~\ref{fig:TAER}).
The Auger measurement shows two distinct elongation rates values, one below and another one above the ankle.
Data from TA instead can be explained with one straight line, consistent with the predictions for proton primaries at all energies.

An important difference to keep in mind between the two data sets is the treatment of the detector resolution.
The measurement from Auger has been unfolded, and can be compared directly with theoretical predictions.
Instead, the simulations shown in Fig.~\ref{fig:TAER} have been smeared using the TA detector biases.
Even though the experimental results cannot be directly compared,
the measured values of $\left<X_{max}\right>$ are within uncertainties, except above $\sim3\times10^{19}$~eV.
The Auger Collaboration also measures the width of the $X_{max}$ distribution (shown on the right panel of Fig.~\ref{fig:AugerER}), 
which represents the shower-to-shower fluctuations.
Air showers initiated by lighter primaries have larger fluctuations than showers produced by heavier nuclei of the same energy.
The observed narrow distribution at the highest energies thus argues for a composition dominated by heavy nuclei.
The TA Collaboration compares the distribution of $X_{max}$ in each energy bin with the predictions for proton and iron primaries. 
The results of the Kolmogorov-Smirnov tests are shown on the right panel of Fig.~\ref{fig:TAER}.
It is clear that above  $\sim3\times10^{19}$~eV there is no enough statistical power to differentiate between light and heavy primaries.

There is another important caveat to take into account when comparing these results with the corresponding model predictions for different primary particles.
The first interactions at the top of the atmosphere occur at energies that are a few orders of magnitude higher than any available measurement from collider experiments.


Assuming proton primaries dominate at $10^{18}$~eV, the Pierre Auger collaboration has recently measured the proton-air inelastic production cross section~\cite{Ulrich}.
Using a Glauber calculation, the inelastic proton-proton cross section can be inferred at a center-of-mass energy of $(57\pm7)$~TeV.
This measurement 
 allows the prediction of the total $p$-$p$ cross section at an energy beyond the reach of the LHC~\cite{Block}.

\begin{figure}[!h]
\centering
\includegraphics[width=1.\linewidth]{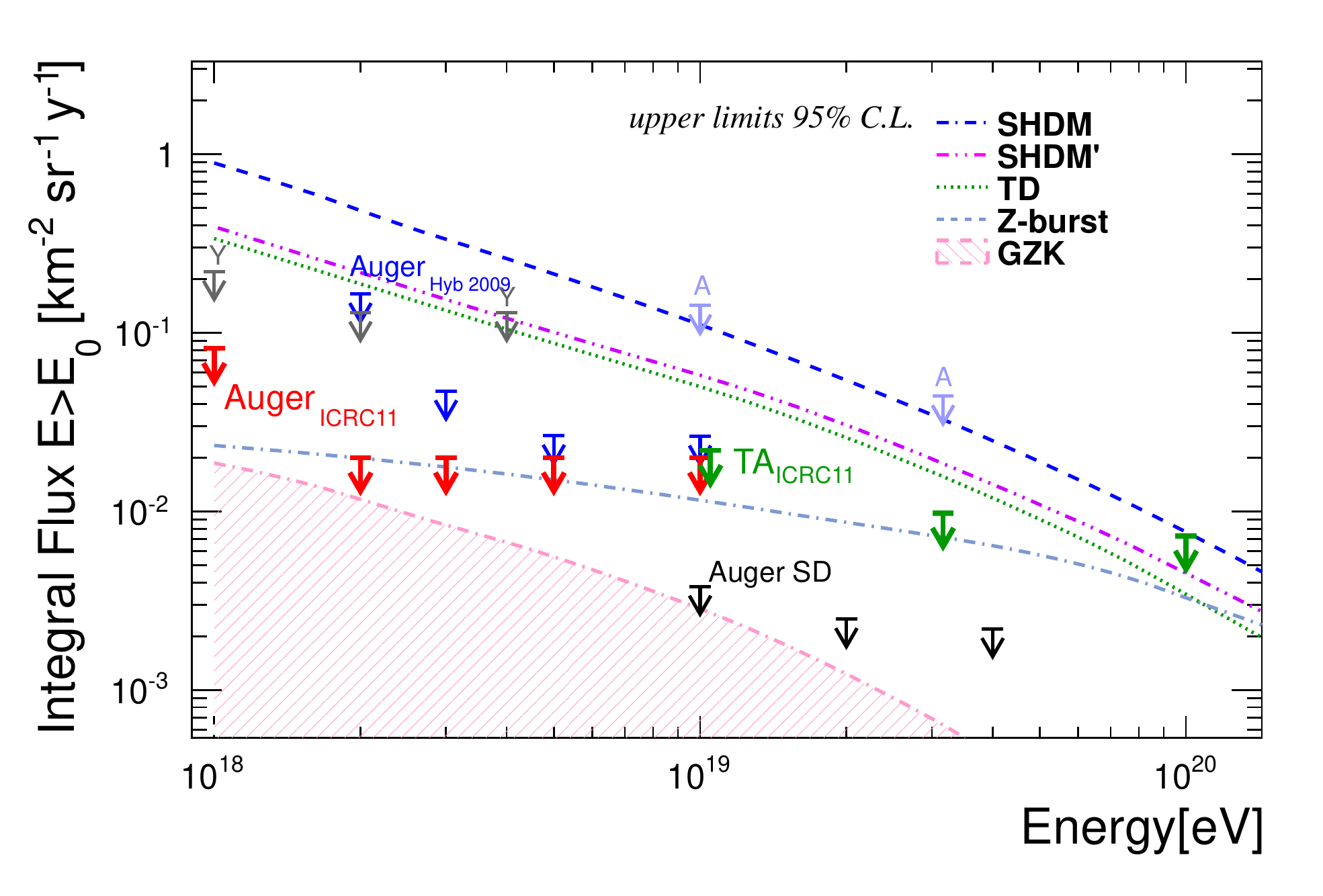}
\caption{Upper limits on the photon flux above 1, 2, 3, 5, and 10~EeV derived from hybrid Auger data~\cite{Settimo} (red arrows),
and above $10^{19}$, $10^{19.5}$, and $10^{20}$~eV derived from TA surface detector data~\cite{Rubtsov} (green arrows)
compared to previous limits.} 
\label{fig:PhotonLimits}
\end{figure}

The measurement of ultra high energy photons is also a crucial probe of the acceleration processes and source models. 
Separating photon-induced from nuclei-induced showers is experimentally  easier than distinguishing between light and heavy nuclear primaries. 
Photon-initiated showers are  driven mostly by electromagnetic interactions, and are therefore less affected by previously described uncertainties in hadronic interactions.
From the properties of the detected showers, both Auger and TA set limits on the fraction of gamma-rays in UHECRs at energies above $10^{18}$~eV.
Current photon limits, shown in Fig.~\ref{fig:PhotonLimits}, have already disfavored most non-acceleration (``exotic'') models of the UHECR origin.

\begin{figure}[!h]
\hspace*{-10pt}
  \begin{minipage}[t]{15pt}
    \vspace{-4pt}\includegraphics[height=0.195\textheight]{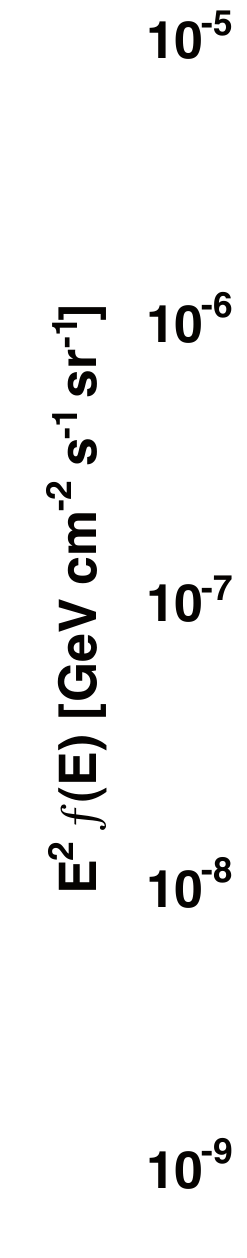}
  \end{minipage}
  \begin{minipage}[t]{210pt}
   \vspace{0pt}\includegraphics[height=0.2\textheight]{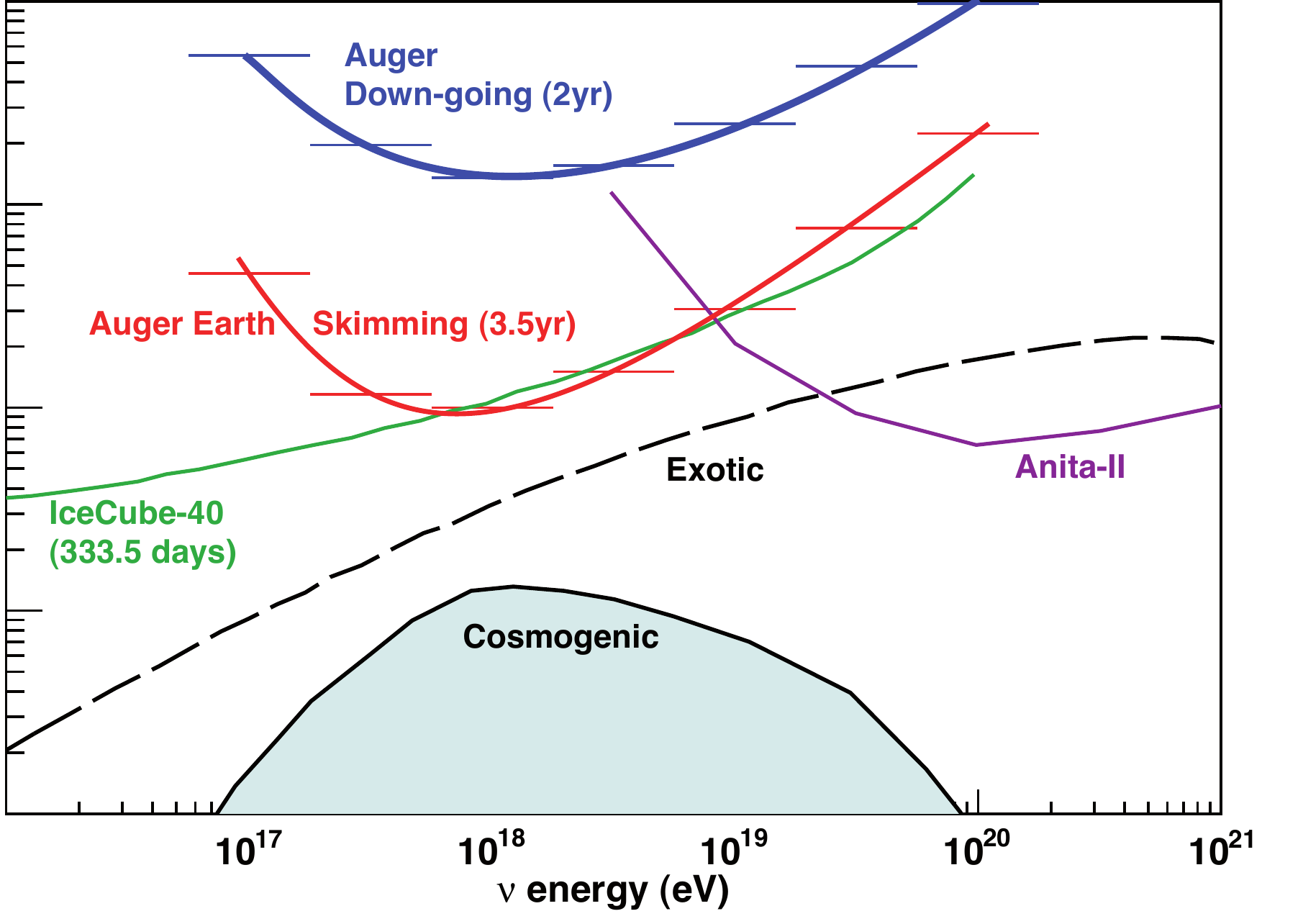}
  \end{minipage}
\caption{Differential upper limits (90\% C.L.) from the Pierre Auger Observatory for a diffuse flux of down-going $\nu$ (2 yr of full Auger), and 
Earth-skimming $\nu_\tau$ (3.5 yr of full Auger)~\cite{Guardincerri}. 
Limits from other experiments are also plotted~\cite{Ref16}. 
Expected fluxes are shown for cosmogenic neutrinos~\cite{Ref17} and for a theoretical exotic model~\cite{Ref18}.} 
\label{fig:NuLimits}
\end{figure}

A large fraction of the energy that protons and heavier nuclei lose in photo-production interactions generates ultra-high energy neutrinos. 
Due to the neutrino small interaction cross section, they suffer only adiabatic energy loss, and we should be able to detect ultra-high energy neutrinos created a long time ago (say, at $z=3$). 
That is why the cosmological evolution of the UHECR sources is a very important parameter in the prediction of the cosmogenic neutrino fluxes.
As mentioned above, the SD of the Auger Observatory can observe air showers with large zenith angles, and therefore it is sensitive to different neutrino channels.
The so-called ``Earth- skimming" tau neutrinos are expected to be observed through the detection of showers induced by the decay products of an emerging $\tau$ lepton, 
after the propagation and interaction of a $\nu_\tau$ inside the Earth. 
``Down-going" neutrinos of all flavours can interact in the atmosphere and induce a shower close to the ground. 
The current Auger limits to ultra-high energy neutrinos (in both channels) are shown in Fig.~\ref{fig:NuLimits}.
Auger has also set competitive limits on ultra-high energy neutrinos from point-like sources in a wide range of declinations~\cite{Guardincerri}. 

\subsection{Anisotropy}

To understand where the UHE cosmic rays come from, one needs to make a careful survey of the arrival directions, and search for both small- and large-scale anisotropies in their distribution.

The large scale distribution of the arrival directions of UHECRs represents one of the main tools for understanding their origin.
The Pierre Auger collaboration has presented an analysis of the first harmonic of the right ascension distribution of cosmic rays using the last seven years of data~\cite{Lyberis}.
The Telescope Array collaboration has also shown the correlation between their first three years of surface detector data and the large-scale
structure of the Universe~\cite{Tinyakov}.
Both analysis show consistency with an isotropic distribution above $10^{18}$~eV.
Furthermore, the Auger collaboration uses their results to derive upper limits on 
the amplitude of the first harmonic and
the equatorial dipole component as a function of energy~\cite{Lyberis},
and also on the flux of neutrons from Galactic sources at EeV energies~\cite{Rouille}.

\begin{figure}[!ht]
\hspace*{-0.5cm}
\includegraphics[width=0.45\textwidth]{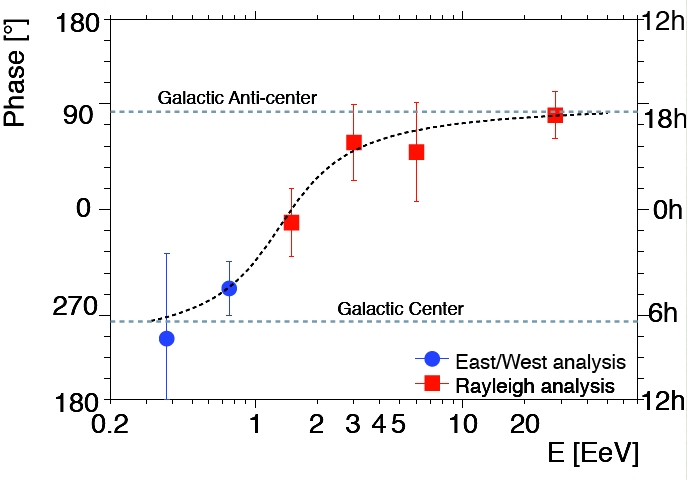}%
\caption{Phase of the first harmonic in the right ascension distribution of cosmic rays detected with the surface detector of the Pierre Auger Observatory as a function of energy~\cite{Lyberis}.} 
\label{fig:LargePhase}
\end{figure}

A very interesting result was found by the Auger collaboration on the phase of the first harmonic of the right ascension distribution, 
shown in Fig.~\ref{fig:LargePhase}, as a function of the energy~\cite{Lyberis}.
While the measurement of the amplitude does not provide evidence for anisotropy, 
the measurement of the phase suggests a smooth transition at around $10^{18}$~eV.
No confidence level could be derived from this result because the study was performed a posteriori.

\begin{figure*}[!ht]
\centerline{\includegraphics[trim=0.1cm 7.1cm 0.2cm 8.5cm, clip=true, width=1.2\textwidth]{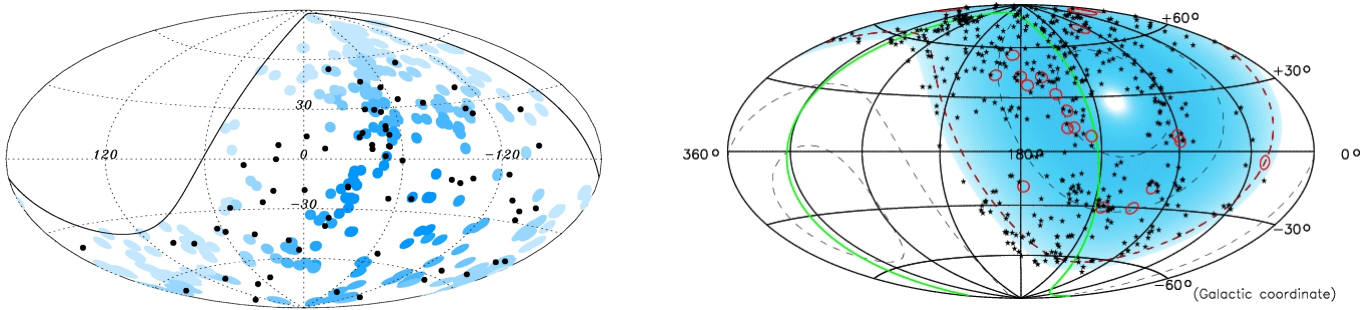}}
\caption{Arrival directions of cosmic rays with energy $E > 55$~EeV detected with the Pierre Auger Observatory (black dots on the left panel)~\cite{correlation},
and with energy $E > 57$~EeV detected with the Telescope Array experiment (red open circles of radius $3.1^\circ$ on the right panel)~\cite{Tkachev},
 in galactic coordinates. 
The border of the field of view of the Auger Observatory (Telescope Array) corresponds to zenith angles smaller than $60^\circ$ ($45^\circ$). 
Blue circles of radius $3.1^\circ$ (Black dots) on the left (right) panel are centered at the positions of the 318 (295) AGNs in the VCV catalog that lie within 75 Mpc, 
and that are within the field of view of each detector.
} 
\label{fig:AGNs}
\end{figure*}

The Pierre Auger collaboration has reported the correlation between UHECRs with energy higher than $57$~EeV and
nearby (closer than 75~Mpc) AGNs from the Veron-Cetty \& Veron (VCV) catalog~\cite{VCV} at an angular scale of $3.1^\circ$~\cite{correlation}. 
The current fraction of correlating events is 33\% (28 events correlate out of a total of 84 detected after their prescription)~\cite{Kampert}.
The TA collaboration has tested this correlation with their SD data (that covers the Northern Hemisphere).
The fraction of TA events that correlate is 40\% (8 out of 20)~\cite{Tkachev}. 
The maps of the events above $5.5\times 10^{19}$~eV and AGNs on both hemispheres are shown in Fig.~\ref{fig:AGNs}.

\section{Conclusions and Outlook}
The small flux of UHECRs requires large detection areas.
There are two experiments, Auger and TA, taking data at these energies.
Having two independent experiments provides not only the necessary cross-checks, 
but also data over the whole sky.
The latest results represent a large amount of high quality data over both hemispheres.
The exposure of the Auger (TA) SD is $21000$~km$^2$~sr~yr ($2700$~km$^2$~sr~yr).

It is important to highlight that, despite the difference in energy scale (of $\sim20\%$),
the features measured by the Pierre Auger and Telescope Array collaborations in the energy spectrum of UHECRs are in very good agreement.
In this regard, I doubt that larger statistics will contribute to a discovery. 
Instead, I think that
the combination of spectral analyses, composition measurements, and small scale anisotropy will be more fruitful in 
our search to identify the sources of UHECRs.

The spectrum of UHECRs is characterized by a flux suppression at energies above $4\times10^{19}$~eV.
The search for this feature has conclusive results mainly due to the statistically significant data collected by the Telescope Array and the Pierre Auger Observatory. 
The spectrum measured by TA (Auger) shows a flux suppression 
(spectral index that changes from $2.68$ to $4.2$)
at the $4\sigma$ ($20\sigma$)  level.
Even though the energy of the suppression matches the expected value due to photopion production during the propagation of protons on cosmological scales,
the possibility that the observed softening is due to the maximum energy of acceleration at the source cannot be completely  ruled out yet.

To understand the origin of UHECRs, it is crucial to infer which cosmic rays are in fact extragalactic. 
There are two main lines of thought in this respect.
One puts the transition above $10^{18}$~eV where a steep galactic spectrum encounters the flat spectrum of extragalactic cosmic rays. 
In the other scenario, the transition takes place at one order of magnitude lower energy.
The two models differ the most from the point of view of the chemical composition. 
If galactic cosmic rays extended to above $10^{18}$~eV, they would be mainly iron nuclei.
But if cosmic rays with energy above $10^{18}$~eV were mostly protons, then the proton-dominated extragalactic component would be important down to even lower energies.
Unfortunately the measured
energy spectra can be explained by very different models of the galactic to extragalactic transition, different injected chemical compositions, and spectral indices~\cite{Kumiko}.
On the other hand, the two collaborations differ the most in the interpretation of their composition-related measurements.

The elongation rate measurements from Auger suggest different composition trends below and above the ankle,
while the TA data are consistent with proton primaries at all energies.
The disagreement between the two measurements is larger above $3\times 10^{19}$~eV,
where the Auger results argues for heavier nuclei, and the TA sample lacks the statistical power.
It is important to remember that both experiments see a suppression in the flux at these energies,
and --due to the telescope's duty cycle and the tight event selection-- 
the data samples are of the order of 5\% the statistics of the respective surface arrays.
It is also possible that the Auger result is actually telling us that there are changes in the hadronic interactions with respect to the current extrapolations from TeV energies,
and not that there is a transition to heavier composition at the highest energies.

The Pierre Auger collaboration recently reported the measurement of the inelastic proton-air section,
and the inelastic $p$-$p$  cross section at $57$~TeV.
These measurements allow the determination of  the inelastic and total cross sections for $pp$ and $\bar{p}p$ interactions at an energy beyond the reach of any man-made accelerator.

Neutrinos and photons are produced by UHECRs when they interact inside their source region or during their propagation to Earth,
and then they will pinpoint back  the position of their production.
Current limits on photons and neutrinos at ultra-high energies have not only ruled out most exotic models for the origin of UHECRs,
but they are also at the verge of testing the most optimistic models for the production of cosmological photons and neutrinos.
The detection of ultra-high energy gamma rays and neutrinos is directly related to our search for the sources of UHECRs, and it
will help us determine (or constrain) the source evolution, the galactic to extragalactic transition, the injected chemical composition, etc.
In particular, the detection of cosmogenic neutrinos could help us understand the origin and type of UHECR 
as their flux depends on the cosmic ray spectrum and composition. 
Neutrinos could also point at cosmic ray sources, and Auger is sensitive to point-like sources of ultra-high energy neutrinos over a large range of declination.

If UHECRs come from a large number of nearby sources that  trace the inhomogeneous distribution of local matter,
the measurement of a large scale anisotropy at Earth provides information on the nature and size of UHECR deflections. 
These deflections are determined by 
the charge of the UHECR   and the intervening magnetic fields.
None of the large scale studies done by the two collaborations have provided so far any evidence for anisotropy above $10^{18}$~eV.
For example, if there is a dipole, the amplitude cannot be larger than $\sim 1\%$.
There is though a very interesting smooth transition in the phase of the first harmonic of the right ascension distribution in the Auger SD data at $10^{18}$~eV.
An independent data set is needed to confirm this result.
Low energy events from the Auger infill array~\cite{Maris} will be crucial for these studies because the SD array is fully efficient above $3\times10^{18}$~eV.
Continued scrutiny of the large scale distribution of arrival directions of cosmic rays as a function of the energy is important to constrain different models for the cosmic rays origin.

The correlation at small angular scales between the arrival directions of cosmic rays of the highest energies and the location of nearby AGNs has remained constant 
(at a correlating fraction of $\sim40\%$) in both data samples over the last three years. 
Between the two experiments there are over 100 cosmic rays with energies above $5.5\times10^{19}$~eV covering the whole sky.

Even though there has been an impressive amount of new results 
in the last year,
we cannot say that we have identified the sources of UHECRs yet.
In the next few years, I expect a significant improvement in the statistics, and in our understanding of the systematic errors and the detection biases.
%
It is also important to mention that both collaborations 
are currently pursuing new detection techniques~\cite{radio, radar}, 
and are also upgrading their detectors to cover the energy range below $10^{18}$~eV~\cite{AMIGA,TALE}.


\bigskip 
\begin{acknowledgments}
This material is based upon work supported by the National Science Foundation under Grant No.~0838088.
Any interpretations of the results, opinions, conclusions, or recommendations expressed in this paper 
are those of the author and do not necessarily reflect the views of the
Pierre Auger and Telescope Array collaborations.
\end{acknowledgments}

\bigskip 
\bibliography{basename of .bib file}

\end{document}